\documentclass[manuscript,screen]{acmart}

\AtBeginDocument{%
  \providecommand\BibTeX{{%
    \normalfont B\kern-0.5em{\scshape i\kern-0.25em b}\kern-0.8em\TeX}}}

\setcopyright{rightsretained}
\copyrightyear{}
\acmYear{}
\acmDOI{}

\acmConference[]{}{}
\acmBooktitle{}
\acmPrice{}
\acmISBN{}

\usepackage{subcaption}

\begin{document}

\title[Retrospective End-User Walkthrough]{Retrospective End-User Walkthrough: A Method for Assessing How People Combine Multiple AI Models in Decision-Making Systems}

\author{Vagner Figueredo de Santana}
\email{vsantana@ibm.com}
\orcid{0000-0003-0325-1596}
\affiliation{%
  \institution{IBM Research}
  \country{United States}
}

\author{Larissa Monteiro Da Fonseca Galeno}
\email{larissagaleno@ibm.com}
\affiliation{%
  \institution{IBM Research}
  \country{Brazil}
}

\author{Emilio Vital Brazil}
\email{evital@br.ibm.com}
\affiliation{%
  \institution{IBM Research}
  \country{Brazil}
}

\author{Aliza Heching}
\email{ahechi@us.ibm.com}
\affiliation{%
  \institution{IBM Research}
  \country{United States}
}

\author{Renato Cerqueira}
\email{rcerq@br.ibm.com}
\affiliation{%
  \institution{IBM Research}
  \country{Brazil}
}

\renewcommand{\shortauthors}{Santana, et al.}

\begin{abstract}
Evaluating human-AI decision-making systems is an emerging challenge as new ways of combining multiple AI models towards a specific goal are proposed every day.
As humans interact with AI in decision-making systems, multiple factors may be present in a task including trust, interpretability, and explainability, amongst others.
In this context, this work proposes a retrospective method to support a more holistic understanding of how people interact with and connect multiple AI models and combine multiple outputs in human-AI decision-making systems.
The method consists of employing a retrospective end-user walkthrough with the objective of providing support to HCI practitioners so that they may gain an understanding of the 
higher order cognitive processes in place and the role that AI model outputs play in human-AI decision-making.
The method was qualitatively assessed with 29 participants (four participants in a pilot phase; 25 participants in the main user study) interacting with a human-AI decision-making system in the context of financial decision-making.
The system combines visual analytics, three AI models for revenue prediction, AI-supported analogues analysis, and hypothesis testing using external news and natural language processing to provide multiple means for comparing companies.
Beyond traditional results on tasks and usability problems, outcomes presented suggest that the method is promising in highlighting \emph{why}  AI models are ignored, used, or trusted, and how future interactions are planned. It also provides a means for eliciting information architecture requirements such as \emph{where} AI model outputs could be placed to increase adoption of AI-supported approaches. For instance, merging recurrent and more traditional tasks (e.g., visual analytics) with newly added models providing recommendations/predictions related to traditional tasks.
We suggest that HCI practitioners researching human-AI interaction can benefit by adding this step to user studies in a debriefing stage as a retrospective Thinking-Aloud protocol would be applied, but with emphasis on revisiting tasks and understanding \textit{why} participants ignored or connected predictions while performing a task.
\end{abstract}



\maketitle

\section{Introduction}

Human-AI decision-making can be defined as the paradigm in which human decision-makers are supported by one or more Artificial Intelligence (AI) models providing recommendations/predictions for a certain task~\cite{Lai2021}.
In recent years, with the advances in AI and its pervasive use in multiple domains (e.g., healthcare, finance, justice, and recruiting), explainable AI (XAI) has become fundamental for humans to assess and make decisions based on  AI models, especially for legal reasons~\cite{Hind2019, Weld2019}.
XAI can be defined as the field that aims to make the output of AI systems more understandable to humans~\cite{Adadi2018}. According to DARPA (Defense Advanced Research Projects Agency), XAI aims at producing more explainable models, while maintaining a high level of prediction accuracy and enabling users to understand, appropriately trust, and effectively manage human AI partnership\footnote{https://www.darpa.mil/program/explainable-artificial-intelligence}. The explanation provided in XAI considers two main characteristics: complexity and domain. The complexity characteristic refers to matching the content (explanation) with the content consumer (user). The domain characteristic highlights the need to tailor the explanation to the domain in which the AI system is being used~\cite{Hind2019}. According to Wang et al.~\cite{Wang2021}, explanations should satisfy three properties: (1)  Improve understanding of the AI model; (2) Help identify model uncertainty; (3) Support calibrated trust in the model.
These works highlight the strong interconnectedness between XAI and trust when AI models are used in decision-making systems.

As new requirements for Human-AI decision-making systems emerge (XAI, trust, uncertainty recognition, and planning future interactions), new methods and tools are needed to help HCI practitioners understand how to bridge the gap between humans and AI decision-making systems to realize the best of this human-AI partnership. However, in contrast to traditional HCI approaches, there is a lack of methods to specifically assess the interplay between different AI models \cite{Lai2021}, especially when recency bias may influence how people deal with recommendations \cite{Fudenberg2016, Chen2019}.

Decision-making is known to be influenced by cognitive processes and affective processes \cite{Fudenberg2016, Shiv1999}. 
Two of the most commonly used techniques that support gaining insights about these processes are Thinking-Aloud Protocol and Cognitive Walkthrough.
Thinking-Aloud Protocol is a technique where participants are encouraged to verbalize their thoughts related to performing tasks while interacting with a user interface (UI) \cite{Lewis1982}.
This technique is commonly employed in formative user studies, supporting the understanding of the rationale and task solving strategies used by participants~\cite{Lazar2010}. Moreover, it has two important variations: Concurrent Thinking-Aloud (CTA), where participants verbalize their thoughts while performing a task, and Retrospective Thinking-Aloud (RTA), where participants verbalize their thoughts while reviewing/replaying their tasks.
However, performing CTA in certain tasks can be intrusive, impact the overall user experience, and influence results \cite{Mandryk2004, Lazar2010, Barbosa2021}. On the other hand, post-test questionnaires may fail to capture nuances of the overall interaction \cite{Mandryk2004, Lazar2010, Barbosa2021}. This complexity motivates the incorporation of the "walkthrough" to bridge this gap, using the system as an artifact to reduce cognitive load as an external representation \cite{Barbosa2021}.
Cognitive Walkthrough (CW) is a review method in which an HCI expert simulates performing tasks to provide insight into how users may interact with the UI~\cite{Lazar2010}. 
In specific decision-making domains, subject matter experts (SMEs) play a key role.  Hence, the method proposed in this work aims at incorporating this aspect by combining RTA with an SME-performed walkthrough in a debriefing phase.
According to \cite{Lazar2010}, debriefing provides a valuable opportunity for sharing thoughts that did not fit into the session, offering an opportunity for participants to share opinions and correct misunderstandings.
 
\subsection{Contribution Statement}

This work is part of a long-term project on human-AI decision-making that explores how SMEs combine multiple AI models in different domains including data science, discovery of new materials, and scientific workflows. While the project considers multiple domains, the user study and the system described in this paper relate to the finance domain. 
This work proposes a method to help HCI practitioners gain a better understanding of how people combine multiple AI model outputs, in a debriefing phase, given that verbalizing all the rationale during the interaction with multiple AI models can be overwhelming for users.
The main contribution of this work is in providing a method, called \textbf{Retrospective End-User Walkthrough}, for screening the level of understanding and composition of AI model outputs, including why certain AI models where ignored by users.
We explore the role of AI model outputs in decision-making, XAI, trust in those outputs, and how SMEs plan future use and interactions with the system after their first use of the system. The method was evaluated with 29 participants; four people participated in a pilot to validate the protocol and identify main usability issues; 25 participants ($n$=25) explored the insights derived by the proposed method.

The remainder of this paper is structured as follows: section \ref{sec:relatedWork} presents related work, section \ref{sec:evaluation} details how the approach was evaluated, section \ref{sec:results} details the results, section \ref{sec:discussion} discusses results obtained so far, and section \ref{sec:conclusion} presents some reflections regarding suitability of the proposed method and our future research directions.

\section{Related Work}
\label{sec:relatedWork}

Thinking-Aloud Protocol is a commonly used technique, present in multiple user studies' protocols. However, Concurrent Think-Aloud (CTA) and Retrospective Think-Aloud (RTA) variations have pros and cons, depending on the study goal and task characteristics. For example, 47\% of the time \cite{Guan2006}, CTA participants are unable to describe areas they look at while interacting with a computer. However, RTA relies on participants' long-term memory to discuss what and why specific actions were taken.
The literature also presents a comparison between CTA and RTA in terms of usability problems and ways to identify them. According to \cite{Haak2003}, both methods reveal comparable usability issues but in different ways. CTA exposes usability problems by means of observed interaction, whereas RTA provides insights via verbalization. Moreover, the authors highlight that CTA may negatively impact task performance and that RTA may be useful in cases of high task complexity. This last aspect aligns with our proposal, given that complex tasks may arise in human-AI decision-making systems, especially when combining multiple AI model outputs. In a study comparing CTA and RTA in the context of health care,~\cite{Peute2015} show that RTA may be valuable when participants' expertise and experience play a key role in the system and on identifying design requirements, verbalizing and reflecting on actions performed, rationales, and describing desired information/features. These characteristics connected to higher order cognitive processes also lead us to think about ways to deal with such challenge in terms of expanding this retrospective step to use the multiple AI model outputs themselves, revisiting how they were part of the cognitive and decision-making processes, as proposed in Distributed Cognition Theory \cite{Hutchins1996, Hutchins2000, Barbosa2021}\footnote{
Distributed Cognition is a theory that encompasses a broad range of cognitive events not restricted by the skin or skull of an individual, rebuilding cognitive science from outside in \cite{Hutchins2000}. In this theory, cognitive activity is considered to be situated in a sociocultural systems and materialized environment, beyond the individual, and distributed in time.
It also describes the human cognition in terms of \textit{"how information processing is dispersed across people and their workplace, their technologies, and their social organization and how information processing evolves over time."}~\cite{Hutchins2000}.}.

Cognitive Walkthrough (CW) was first proposed in \cite{Lewis1990} as a means to systematically evaluate features of an interface in the context of cognitive learning theory and how users create a mental model of the UI. CW allows for evaluating how a user with minimal system knowledge performs a task \cite{Mahatody2010}. The authors indicate that CW, beyond simple, could be difficult to apply depending on the specific study context \cite{Mahatody2010}.
The literature discusses numerous variations of the original CW method as  
Heuristic Walkthrough \cite{Sears1997}, 
Norman’s CW \cite{Rizzo1997}, 
Streamlined CW \cite{Spencer2000}, 
CW for the Web \cite{Blackmon2002}, 
Groupware Walkthrough \cite{Pinelle2002}, 
Activity Walkthrough \cite{Bertelsen2004}, 
Interaction Walkthrough \cite{Ryu2004}, 
CW with Users \cite{Granollers2006}, 
Distributed CW \cite{Eden2007}, and
Cognitive Barriers Walkthrough \cite{Santos2019}.
The most relevant for the current proposal is \textbf{CW with Users} \cite{Granollers2006}; for a thorough review of CW, please refer to \cite{Mahatody2010}. 

CW with Users explicitly integrates users into the walkthrough process. It is comprised of three phases~\cite{Granollers2006}: In Phase 1, CW is performed in the traditional manner. In Phase 2, representative system users are incorporated into the process. After a brief introduction, they are invited to perform all the tasks defined in Phase 1 that correspond to their profile. During this interaction, the users are asked to use CTA to express their thoughts, feelings, and opinions. Users perform the tasks without any explanation beyond the brief introduction. At the end of each task, they note the main deficiencies detected. Once the users have completed the tasks, they are invited to comment on any problem identified during Phase 1. In Phase 3, experts review the concerns identified by users during Phase 2. 

In recent works, \cite{Putri2021} presents an evaluation of a restaurant finder using CW. The evaluation consisted of an enhanced CW, User Experience Questionnaire (UEQ) and System Usability Scale (SUS). As a result, the authors present that they were able to identify significant problems that became improvement suggestions. 
In \cite{Schoeffer2022}, the authors present that different amounts of information and self-assessed AI literacy influence how people perceive fairness in decision-making system support.
In \cite{Alomari2020}, a method is proposed for evaluating a cyberlearning environments interface. They use techniques such as CW with CTA and a heuristic evaluation survey and apply it to specific software, SEP-CyLE. Our work  also proposes a method for evaluating interfaces using CW and CTA and testing it with specific software. However, we propose an evaluation method for knowledge-centric systems for understanding how people create insights from (or ignore) multiple AI model outputs. We have a debriefing phase at the end of the session, allowing the user to perform an RTA while navigating the interface and reasoning and detailing rationales about choices made.  

The main difference between the presented works and our proposal is in making the process more straightforward, bearing in mind the structured debriefing around human-AI interaction (e.g., knowledge-creation, XAI, trust, and future planned interactions) and RTA allowing participants to recall their actions while navigating the UI. The goal of this proposal is to make the method easier to learn and apply and less laborious to collect and analyze data. Moreover, when comparing our proposal against traditional CW-based methods, a key aspect of our approach is identifying how people connect multiple AI model outputs, detailing the interaction and knowledge creation around human-AI interaction. CW-based traditional methods aim at identifying usability problems and they are compared in terms of effectiveness in finding these problems. In our method, the goal is to assess how people connect multiple AI model outputs and also on providing insights on how to connect new AI models with existing, ``traditional'' tasks. This aspect grounds our experiment design on comparing the group of people interacting/mentioning AI versus the group of people that did not interact with/mention AI models while performing the tasks/debriefing.

\section{Evaluation of the Proposed Method}
\label{sec:evaluation}

In this section, we present the Retrospective End-user Walkthrough method and how it was empirically assessed with the 25 participants. The proposed \textbf{Retrospective End-user Walkthrough} is detailed in Table \ref{tab:rcw}. This user study was submitted to our 
Healthcare Research Studies Review Board (HRRB)\footnote{In our organization, the HRRB is one of the first groups to assess studies involving humans interacting with prototypes we develop. All studies that involve human participants must be reviewed and approved by the HRRB.} for assessment. The study was approved and the Review Board provided instructions for recording video during the remote studies to retain privacy of study participants during the recordings. Additional details are provided in the materials and procedure subsections (Sections \ref{sec:participants} and \ref{sec:materials}) .

\begin{table}
  \caption{Retrospective End-user Walkthrough.}
  \label{tab:rcw}
  \begin{tabular}{p{3.5cm}p{11.5cm}}
    \toprule
    \textbf{First phase (preparation)} & Tasks and study plan are prepared by HCI practitioners and domain experts. \newline \newline Identification of representative tasks is fundamental for study participants to perform meaningful tasks. Identification of representative study participants can be performed as part of broad stakeholder analysis to ensure diverse participation. Stakeholder analysis may be performed with the support from HCI and/or design professionals.\\ \midrule
    \textbf{Second phase (test)} & Representative system users are incorporated into the process. \newline \newline After a brief introduction to the system via video or structured demonstration, participants are invited to perform tasks defined in Phase 1, in the most natural and meaningful way possible. \newline \newline During this interaction, participants may also be asked to use CTA to express their thoughts and opinions. Use of CTA encouraging participants to verbalize their thoughts should be performed carefully, with consideration for cognitive load. For instance, facilitators can gauge higher levels of cognitive load via longer silence intervals or how participants express themselves while trying to make sense of the output of the multiple AI models. Here, CTA can be used as a warm-up for the  debriefing phase, which is core to the proposed method.\\ \midrule
    \textbf{Third phase (debriefing)} & RTA takes place using the system as an artifact for reflecting on tasks performed. \newline \newline This phase is the core step of the proposed method. Here, participants are invited to perform RTA while interacting with the system being evaluated (walkthrough), reflecting on tasks performed, rationale for choices, decisions, and the like. This step is grounded on Distributed Cognition \cite{Hutchins2000} and aims at supporting participants reflect about their decision-making tasks as they revisit the system and 'replay' their cognition processes. During the debriefing, the facilitator asks guiding questions towards system's features and goals. \newline \newline In the context of decision-making systems combining multiple AI models, the following questions were proposed to cover knowledge-creation, XAI, trust, and future planned interactions; guiding questions may be tailored to systems' characteristics \textit{as long as they encourage participants to revisit tasks performed and use the system as an extended artifact to support reflection on cognitive processes}.
    \begin{enumerate}
        \item Why did you use some system features and not others? 
        \begin{itemize}
            \item \textit{Some examples of used/ignored system features may be provided. Facilitator can mention features the participant interacted with and other features presented at the beginning of the session.}
        \end{itemize}
        \item How did you interpret the output of the AI models? 
        \begin{itemize}
            \item \textit{Some examples of AI model outcomes used by participants may be provided. Facilitator can mention some of the results used by participants while interacting with the system.}
        \end{itemize}
        \item What was your level of trust of the AI models you used? Why?
        \item Going forward, would you change any of the strategies you used?
        \begin{itemize}
                \item \textit{The rationale here is to assess how participants summarize their overall understanding of the system, given test and debriefing phases, and express how they plan future interactions. }
            \end{itemize}
    \end{enumerate}
    \\
  \bottomrule
\end{tabular}
\end{table}

In the following sections we discuss evaluation of the proposed method with the aim of forming a foundational understanding of how people interact with multiple AI model outputs to make decisions via an empirical study analyzing human-AI interaction.
The main research question driving this work is the following: 
\begin{itemize}
    \item \textit{How can we identify how people interact with and combine multiple AI models in decision-making systems?}
\end{itemize}

\subsection{Participants}
\label{sec:participants}

Users were invited to participate on a volunteer basis. All volunteer participants were invited via official channels in our organization. 
A message requesting volunteers for a study was broadcast. The message included the requirement that participants have a basic understanding of finance including financial metrics, basic stock analysis, and asset comparison. Potential study participants contacted the study lead researcher and were screened with questions on knowledge of artificial intelligence and finance fundamentals. Participants ranked their knowledge in these areas from 1 (no knowledge) to 7 (expert).

The following mean values show how participants ranked their knowledge of financial metrics, artificial intelligence, and human-AI decision-making in the finance domain:

\begin{itemize}
    \item Knowledge of Artificial Intelligence: 4.17 ($\sigma=1.27$); 
    \item Knowledge of Explainability: 3.74 ($\sigma=1.25$); 
    \item Knowledge of Financial Metrics: 3.87 ($\sigma=1.46$);
    \item Knowledge of Stock Market: 3.96 ($\sigma=1.36$); 
\end{itemize}

That is, study participants had intermediate knowledge of the aspects that the system and user test would cover.

The study was performed individually in a formal, synchronous and remote fashion \cite{Santana2022}, due to COVID-19 pandemic restrictions. Actions were taken to ensure that use of a remote study did not impact the quality or reliability of the results (e.g., physical and virtual environment were setup to prevent interruptions and promote communication between participant and facilitator).
The duration of each session was approximately one hour. 

\subsection{Materials}
\label{sec:materials}

The materials considered in the study include the following: a consent form, a glossary, remote conferencing software, video recording software, the decision-making system, and the User Experience Questionnaire (UEQ). The consent form and a glossary\footnote{
http://(redacted)
} of financial terms used in the system were emailed to the participants one day prior to their session. The remote conferencing software used was (redacted).
Quicktime software was used to perform screen and audio recording, as recommended by HRRB. It enabled recording only the portion of the screen area associated with the screen the participant was sharing, preventing the recording to register participant's face and name.

The system used in this user study is part of a long-term project developed to support asset comparison through a combination of multiple AI models and visual analytics.
Details of the system visual analytics features and AI-models are provided next.

\begin{figure*}
  \centering
  \begin{subfigure}{10cm}
      \includegraphics[width=10cm]{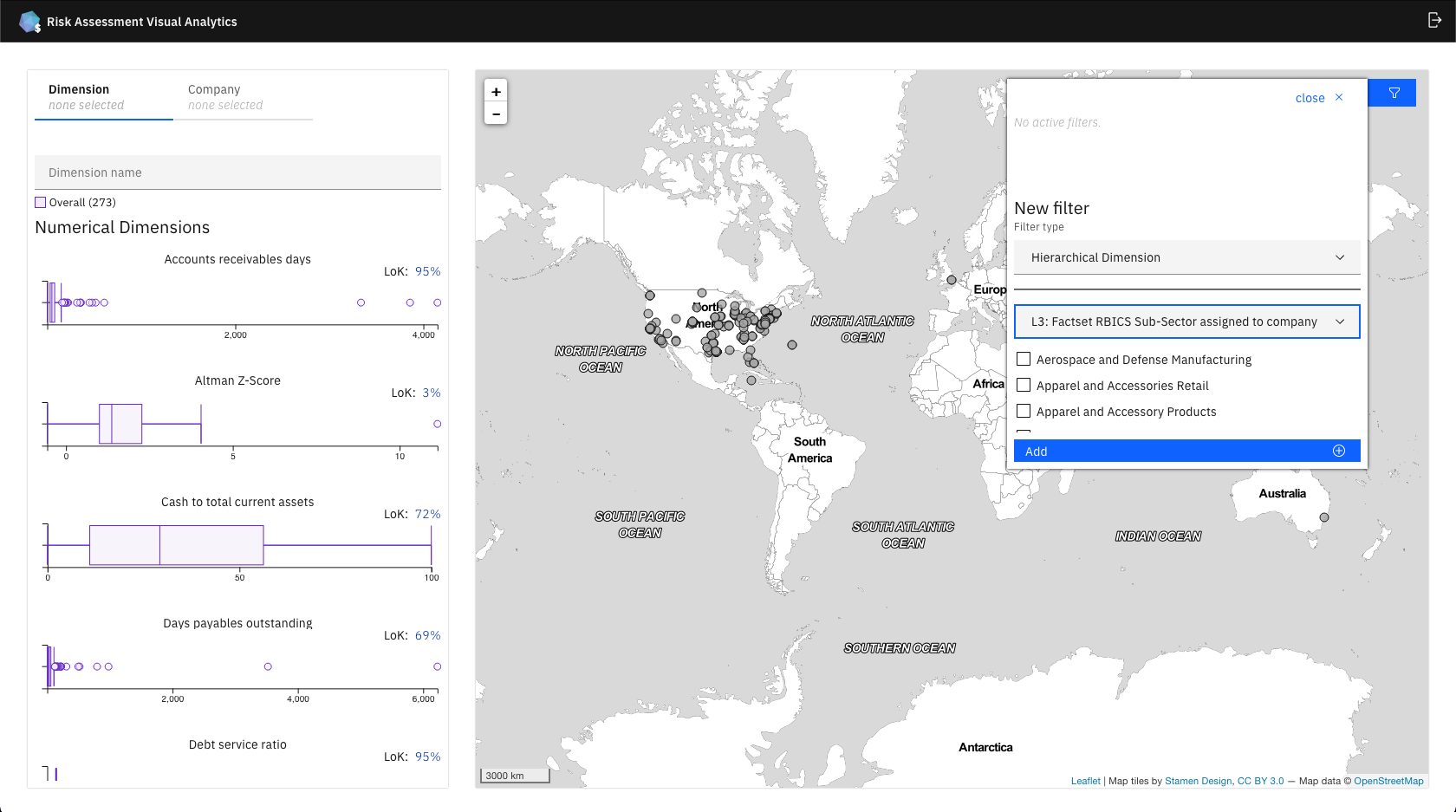}
      \caption{Map and filter features.}
      \Description{Map and filter features.}  
  \end{subfigure}
  \begin{subfigure}{10cm}
      \includegraphics[width=10cm]{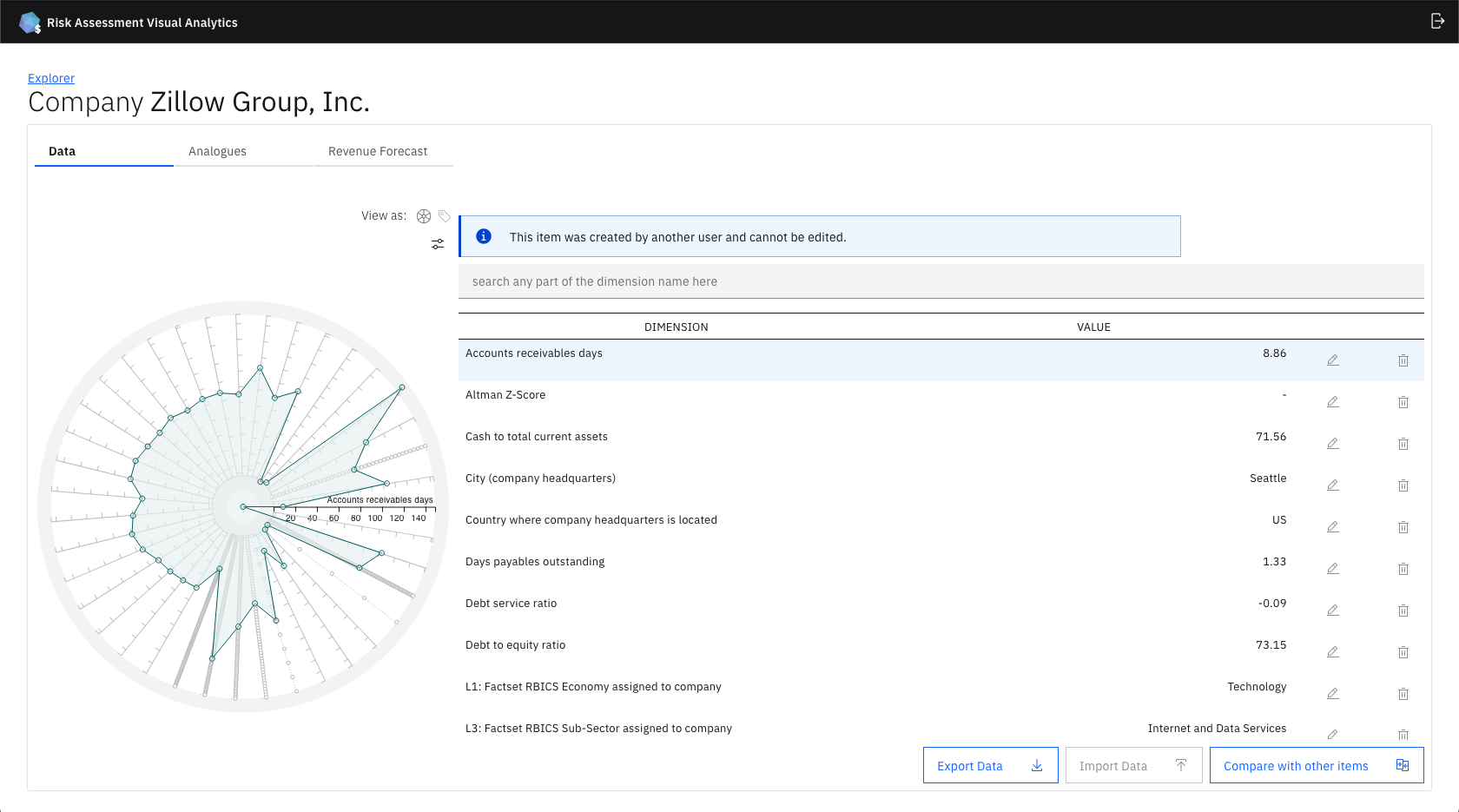}
      \caption{Company details visualization.}
      \Description{Company details visualization.}  
  \end{subfigure}  
  \begin{subfigure}{10cm}
      \includegraphics[width=10cm]{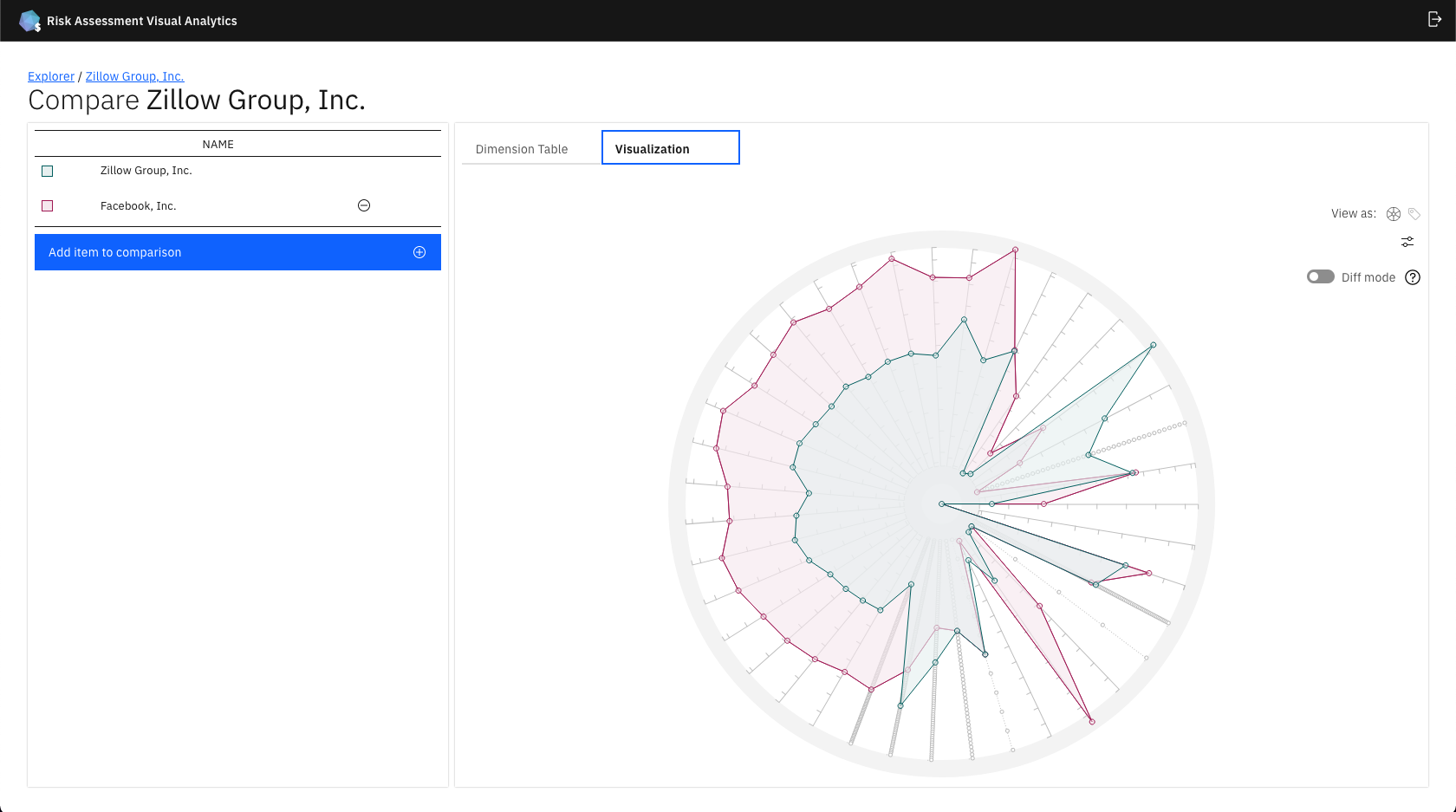}
      \caption{Company comparison visualization.}
      \Description{Company comparison visualization.}  
  \end{subfigure}
  \caption{Visual Analytics provided in the human-AI decision making system considered in the user study.}
  \Description{Visual Analytics provided in the human-AI decision making system considered in the user study.}    
  \label{fig:visual_analytics}
\end{figure*}

The visualizations provided included a map (Figure \ref{fig:visual_analytics}(a), center) showing the location of the companies, represented by points colored according to a selected metric or filter, a list of Box-Whiskers plots (Figure \ref{fig:visual_analytics}(a), left) summarizing distributions of all metrics, company details visualization (Figure \ref{fig:visual_analytics}(b)), and radar plots used to compare companies based on selected attributes (Figure \ref{fig:visual_analytics}(c)). 
Participants could visually inspect the data -- either point in time or historically -- to observe patterns in data over time or to compare performance of companies against each other. The visual inspections could help participant decide which companies may be candidates for investment.
The visualizations provided were interactive and allowed users to select attributes of interest.  Users could also filter data points based on value ranges for all available attributes. The rationale for having an interactive tool was to support the data investigation at multiple levels of detail, as proposed by \cite{Munzner2014}.

\begin{figure*}
  \centering
    \begin{subfigure}{10cm}
      \includegraphics[width=10cm]{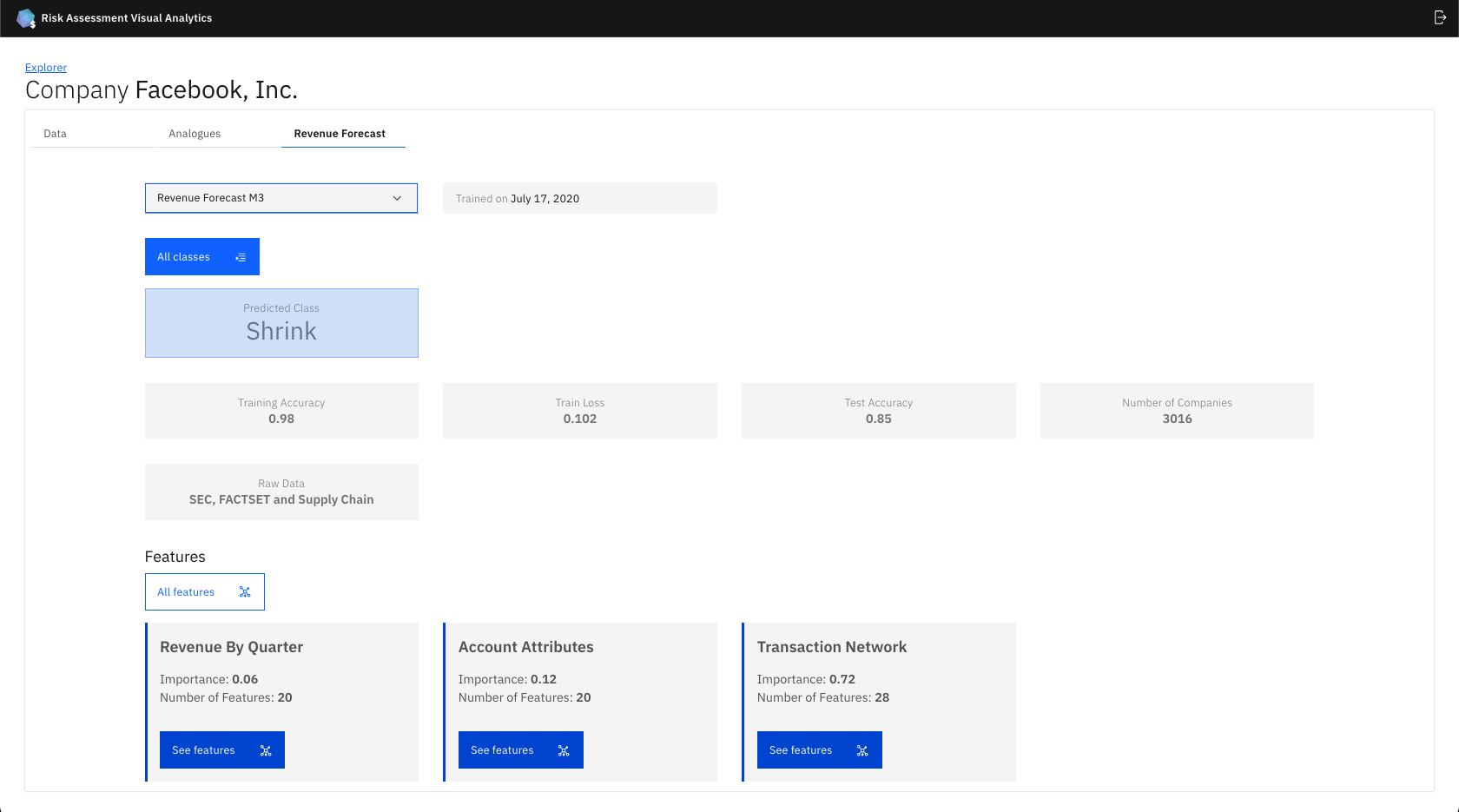}
      \caption{Revenue forecast.}
      \Description{Revenue forecast.}  
  \end{subfigure}
    \begin{subfigure}{10cm}
      \includegraphics[width=10cm]{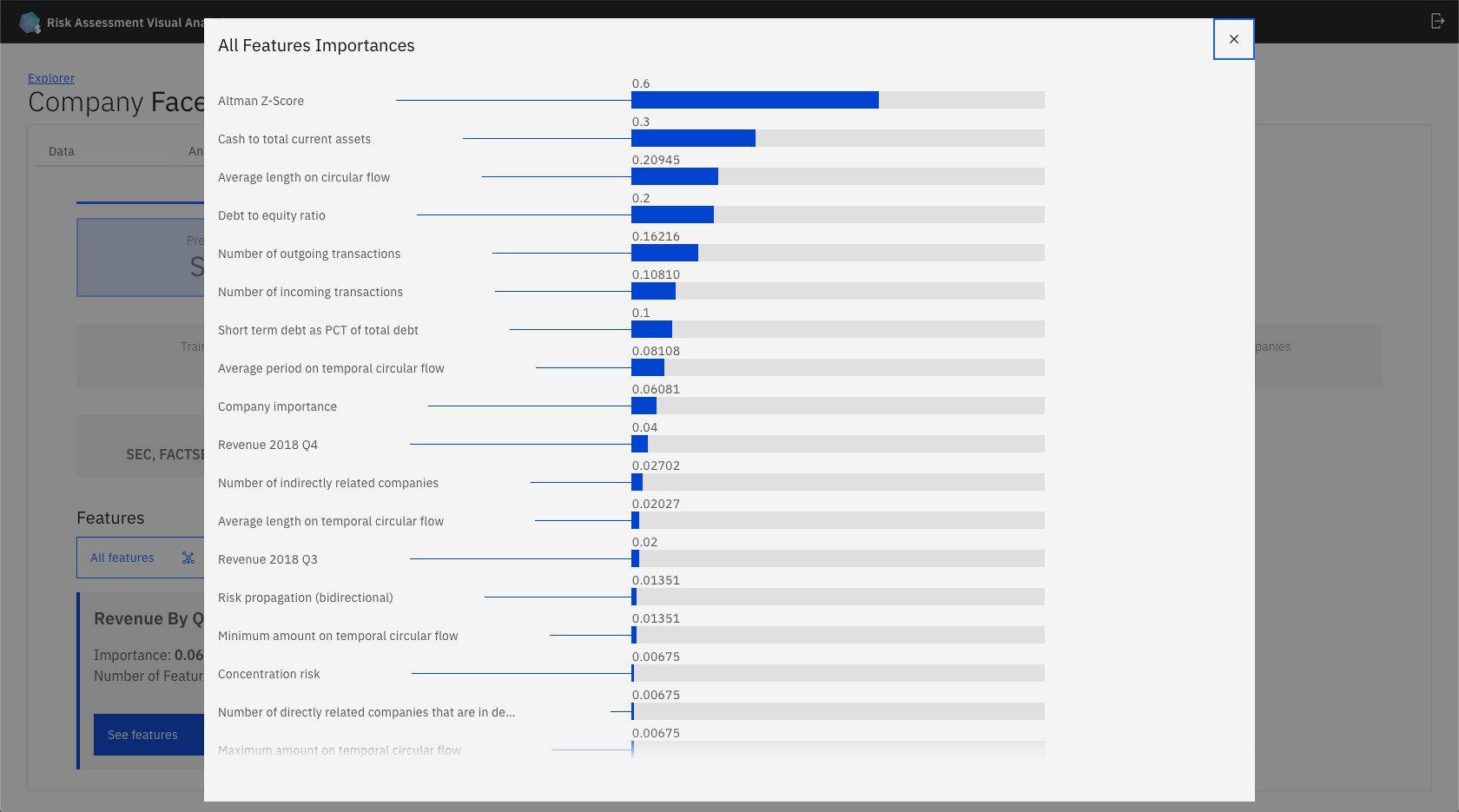}
      \caption{Feature importance for the revenue prediction shown.}
      \Description{Feature importance for the revenue prediction shown.}  
  \end{subfigure}  
  \caption{Revenue prediction model provided in the human-AI decision making system considered in the user study.}
  \Description{Revenue prediction model provided in the human-AI decision making system considered in the user study.}    
  \label{fig:revenue}
\end{figure*}

The revenue prediction feature consisted of three AI models for predicting one of the following classes:  shrink, stable, and grow (Figure \ref{fig:revenue}). The classes were designed to support decision-making when evaluating next quarter revenue. The difference between the models lay in the attributes and datasets used in training the models. The system reported feature importance for the AI models users were interacting with (Figure \ref{fig:system}(b)).

AI-supported analogues analysis aimed at allowing SMEs to analyze sparse, often incomplete datasets to find similar objects. In this system, the objective is to support SMEs to identify similar companies, even when some have missing data. In this case, the SME can select a ML algorithm to infer missing attributes before looking for companies similar to a company of interest. 

\begin{figure}
  \centering
  \includegraphics[width=10cm]{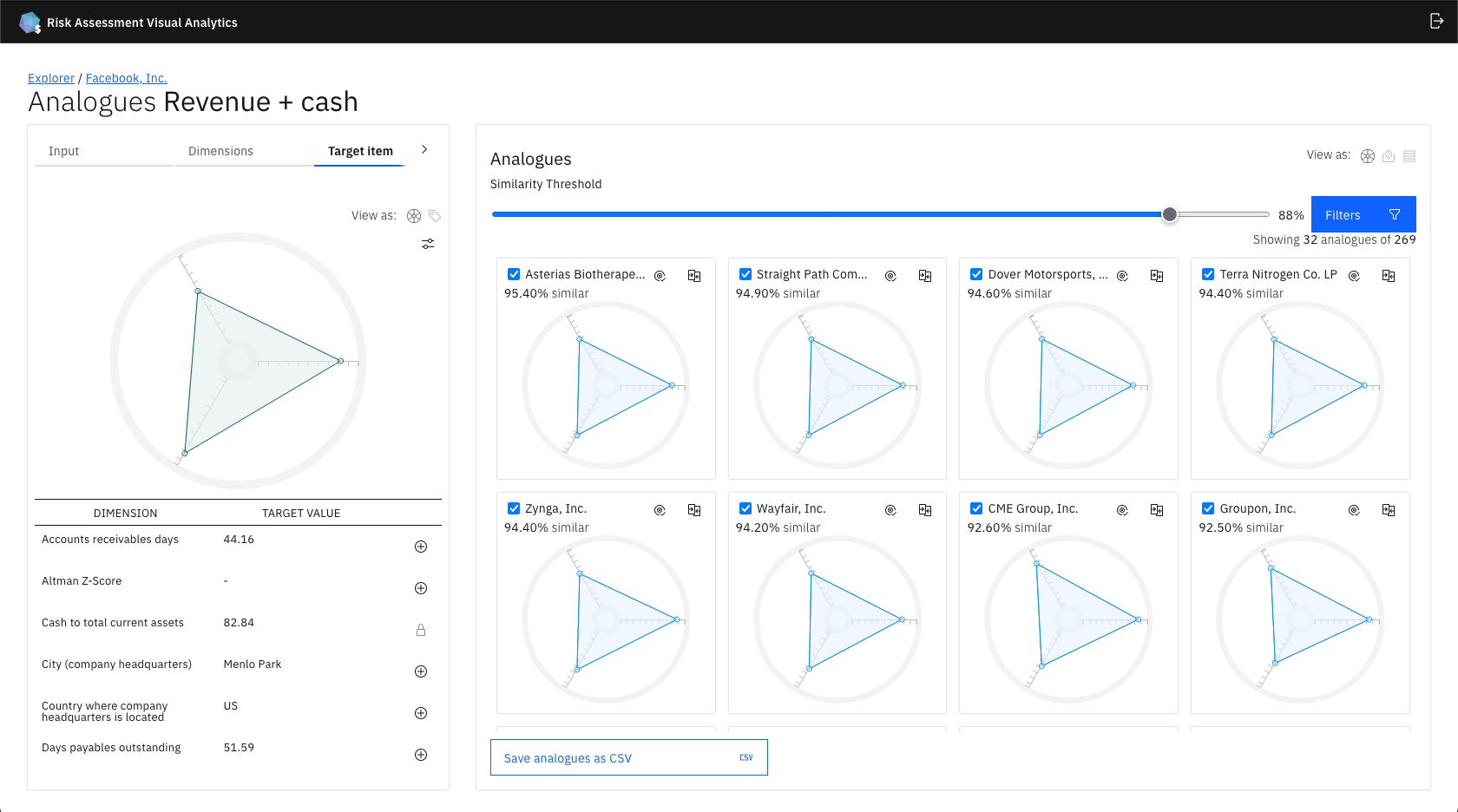}
  \caption{AI-supported analogues analysis.}
  \Description{AI-supported analogues analysis.}    
  \label{fig:analogues}
\end{figure}

\begin{figure}
  \centering
  \includegraphics[width=5cm]{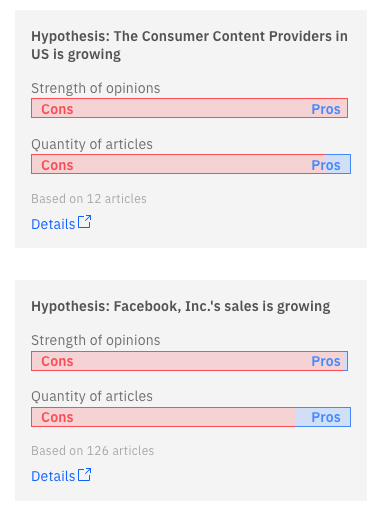}
  \caption{Hypotheses testing using natural language processing and pieces of news.}
  \Description{Hypotheses testing using natural language processing and pieces of news.}    
  \label{fig:hope}
\end{figure}

The hypothesis testing feature provided users external news as a data source and natural language processing to process a set of hypotheses, such as e.g., growth in terms of sales, market share, revenue, or acquisitions (Figure \ref{fig:hope}). With these hypotheses, the system  searched for news about the company of interest (entity analysis) and assessed the news' content to identify whether or not the hypotheses were supported by the assertions found in the text. Then, the results provided to participants show the opinions and quantity of articles pro/con each of the hypothesis (Figure \ref{fig:hope}). 

Tasks that can be performed in the system include revenue prediction analysis, asset comparison, and analogues search, among others. Tasks considered in the study will be detailed in the next section. In addition to the AI models available in the system, the decision-making system was loaded with four years of historical data for each of the 268 companies included in the study. The system was also loaded with historical data relevant to evaluating investment decision-making, including 39 finance metrics relevant for the task.

Finally, to assess participants' perceived User Experience (UX), the User Experience Questionnaire (UEQ)\footnote{https://www.ueq-online.org/} was used as a post-test questionnaire.

\subsection{Experiment Design}

Prior to the main experiment, a pilot study was conducted (participants P1 to P4) to validate the procedure materials and tasks. Domain experts selected for the pilot also provided answers that were used as a benchmark, as detailed in the result analysis section.
Usability problems identified during the pilot were corrected prior to the main experiment (participants P5 to P29).
The experiment design consisted of a brief introduction to the project and the system, followed by a four-minute demo video. Participants  were given five minutes to freely explore the system. After that, tasks were presented to the participants according to the order detailed in Table \ref{tab:task_order}. The rationale for presenting tasks in an alternating sequence was to prevent any learning curve effect.

\begin{table*}[h!]
  \caption{Task order.}
  \label{tab:task_order}
  \begin{tabular}{p{2cm}p{6cm}p{6cm}}
    \toprule
    Participant ID & First task & Second task\\
    \midrule
    Odd & Comparison of two given companies. & Rank 3 out of 8 companies from a given sector.\\
    Even & Rank 3 out of 8 companies from a given sector. & Comparison of two given companies.\\
  \bottomrule
\end{tabular}
\end{table*}

The context presented to participants set a scenario in which they had to select which company's stock should be purchased for investment or should be recommended to a friend or parent. Two different tasks were presented to the users. One task involved choosing to invest between one of two stocks. The second task involved ranking three out of eight stocks from the same sector.  
Tasks were designed with help from a domain expert to support a better understanding of two common scenarios in asset comparison: (1) selection of a company from a reduced set of companies and (2) ranking companies from a given sector (larger set). 
Users were instructed that they could use any of the system's available features to support their decision-making. In addition, given that simulation and real financial decisions differ~\cite{Deniz2020}, participants were presented a gift card reward upon completion of the tasks. The rationale here was to bring some form of real reward for the simulated tasks. Participants were given an unlimited amount of time to complete the tasks. 

Finally, in the debriefing phase, the proposed method was applied to participants who used AI models or mentioned AI models in the debriefing, triggering the Retrospective End-user Walkthrough. This experiment design was employed to allow the comparison between these groups and support the analysis of the insights provided by the proposed methods. In section \ref{sec:resultAnalysis}, we compare insights derived from these two groups.

\subsection{Procedure}

In the pre-test phase, the facilitator verified the participant's understanding of the consent form and participant agreement content.  Three participants asked the facilitator to explain the content of the consent form. Next, the facilitator described the project and context of the study and played the four-minute demo video. Following the video, the facilitator inquired whether the participant had any questions about the features presented in the video or the study context itself. Participants were instructed that they could take notes and were told to explain how/when they took notes and the content of the notes (information need). 
After this initial conversation, participants received a link to log into the system. Immediately upon opening the system's web page, participants were instructed to share their screen via the virtual conferencing software. Video recording began as soon as the screen was shared, capturing only the system area. The participant was given five minutes to freely explore the system.  

In the test phase, the facilitator presented the context of selection/ranking of companies, highlighting that companies could be selected considering best financial health or any set of metrics that the participant deemed relevant. Participants were also informed that there was no time limit for performing tasks and that they could leave the experiment at any time. They were also asked to verbalize and detail any specific metric(s) they considered.
Debriefing took place in the post-test phase, following the method detailed in Table \ref{tab:rcw}. Following the debriefing, participants were asked to provide offline feedback about their perceived UX based on the UEQ.

\subsection{Result Analysis}
\label{sec:resultAnalysis}

In this user study the following data were collected: thinking-aloud data, video interaction, task times, responses to each task, groups of participants that interacted with/mentioned AI models vs. participants who did not interact with/mention AI models, and self-reported UX. Responses to each of the tasks were compared with benchmarks provided from domain experts selected in the pilot. The proposed method was applied considering participants who used AI models or mentioned AI models in the debriefing. Other participants followed the simple debriefing, i.e., without the proposed Retrospective End-User Walkthrough. In the next section, we compare insights yielded by these two groups.


Finally, the analysis aimed at comparing:
\begin{enumerate}
    \item In the test phase, users who interact with AI models vs. users who do not interact with AI models;
    \item In the post-test phase (debriefing), during Retrospective End-User Walkthrough, participants who mention AI models vs. users who do not mention AI models;
    \item Closing the debriefing, users who mention AI models and how they plan combining the knowledge created during the whole session in new ways.
\end{enumerate}

\section{Results}
\label{sec:results}

Task comparison between groups (Table \ref{tab:task_results}) shows that all participants who interacted with AI models got correct answers for the task of comparing two companies, while some participants answered differently (78.6\% for correct answers). On the other hand, average correct answers for the task of ranking 3 out of 8 companies is similar between groups.  In addition, average task time for performing the tasks for the group that interacted with AI models is approximately twice as large as compared with the group that did not interact with AI models. This aspect was  especially pronounced in cases where participants first performed a holistic analysis followed by a more in-depth analysis using AI models to support an initial decision.

\begin{table*}[h!]
  \caption{Task results by group; correct answers consider the expert answers as benchmark.}
  \label{tab:task_order}
  \begin{tabular}{p{3.5cm}|p{2cm}|p{2cm}|p{3cm}|p{2cm}}
    \toprule
    Group & \multicolumn{2}{|c}{Comparison of two companies} & \multicolumn{2}{|c}{Rank 3 out of 8 companies}\\
     & Correct answer & Avg. task time & Avg. correct answers & Avg. task time
    \\ \midrule
    Interacted with AI & 100\% (5/5) & 11:25 & 58.3\% (1.75 out of 3) & 36:24  \\
    Do not Interact with AI & 75\% (15/20) & 05:57 & 59.3\% (1.78 out of 3) & 18:57 \\
  \bottomrule
\end{tabular}
\label{tab:task_results}
\end{table*}



Considering the comparison of participants who interacted with or mentioned AI models vs. participants that did not interact with or mention AI models, five of the 25 participants used AI models to answer or to confirm a choice ("second opinion"). Revenue forecast models were used by five participants; two of them also used the hypotheses testing models based on news.
In addition, of the 20 participants who did not interact with AI models, four participants mentioned in the debriefing that, after starting the task, they forgot to use the additional AI features. P7 said: \emph{"I forgot [the models]"}. P18 reported that: \emph{"I didn't even remember that [Revenue Forecast]"}. P15 and P19 emphasized that they did not use AI models because it was hard to find, respectively: \emph{"I did not know how to return [to the AI model]"} and \emph{I did not use because I did not find it}. These responses highlight that human-AI decision-making systems combining multiple models should balance complexity against number of models used to support user decision-making. Additional feedback related to the inherent complex nature of such systems, P19 reported a click-based criticism: \emph{"It took me 2 clicks to see the level of knowledge but to access the revenue forecast I needed 4 clicks"}. 
This highlights an opportunity for bringing AI model outputs up front or to merge them with more common tasks. 
P14 explained: \emph{"The model takes you to a path, while the visualization gives you an overview"}. And P22 complemented: \emph{"I used the visualization because it is more global"}.

Regarding trust of model outputs, participants mentioned that they trust the system because they know the people responsible for the technology. P12 said: \emph{"I trust because I know who is behind this technology and also due to my background"}. And P19 argued: \emph{"My trust is based on the methodology behind the models"}. These quotes suggest how trust in the system can be expressed in terms of transitivity or similarity (projected) to the ones developing the system. 

Explainability effects also emerged when talking about trust. P18 suggested: \textit{"When selling this, it would be interesting to show how it is done"}. And P20 asked: \emph{"What are the fundamentals of the hypothesis [tested using news processing]?"}; and commented: \emph{"I'm curious to understand the fundamentals used [in the revenue forecast]"}. 

For the group of participants who did not interact with AI models but during the debriefing returned and explored the system and interacted with AI models (walkthrough), two of 14 participants mentioned that they would change their strategies and combine AI models. These two sessions were valuable for understanding why they did not initially use the AI models and, most importantly, how that first interaction could be improved to achieve the full potential of the decision-making system. After interacting with AI models, P7 exclaimed: \emph{"I'm starting to change my opinion!"} and \emph{"I would have to analyze everything again!"}. In addition, P18 mentioned: \emph{"I would use more the revenue forecast"} and \emph{"It seems the best way of doing... analyzing it [revenue forecast] up front"}. These quotes highlight valuable aspects of the Retrospective End-user Walkthrough as proposed, how it can be used to elicit information architecture requirements in decision-making systems, and provide insights on new ways of integrating multiple AI models.

\begin{figure*}
  \centering
  \includegraphics[width=0.7\linewidth]{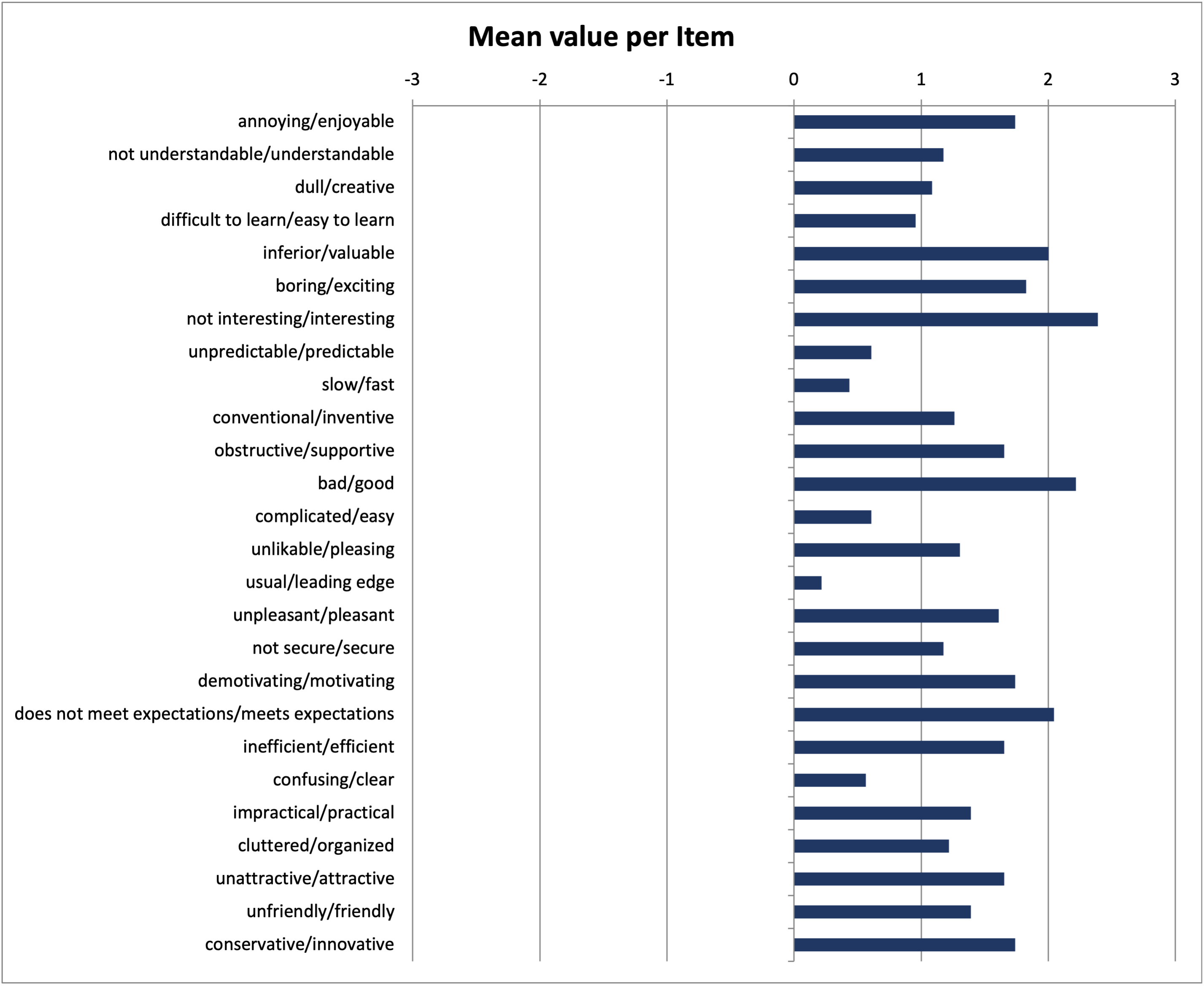}
  \caption{Mean value per UEQ item; negative values represent negative opinions and positive values represent positive opinions provided by participants.}
  \Description{Mean value per UEQ item; negative values represent negative opinions and positive values represent positive opinions provided by participants.}
  \label{fig:ueq}
\end{figure*}

Considering self-reported UX, figure \ref{fig:ueq} presents the mean of UEQ results. It is possible to highlight some relevant aspects of the graph.  For example, the value of "slow/fast" item is 0.1, which means that participants felt that the system was slow. This may have impacted how participants performed their tasks. Participant P18 commented: \emph{"Since this top part [with hypotheses testing] kept loading, I didn't scroll and just went to another page"}. This reveals how loading times plays an interesting role when users have multiple AI models available and how they may switch between available models given the perceived efficiency.  In contrast, participants reported that the system is interesting, being the UEQ item with the highest mean value (2.3). To illustrate that, one participant provided feedback that the software has potential to become a great tool for risk assessment; other participants even asked to be notified as soon as the system is launched. 

\section{Discussion}
\label{sec:discussion}

Based on the results and time to complete tasks, it was possible to identify that people from the group that interacted with AI had better performance in terms of task completion however, when completing the task of comparing two companies, they took approximately twice as long to complete this task as the group that did not interact with AI. This effect can be explained by situations in which participants used AI models to confirm choices made, using the AI models as a second opinion (confirmation bias). Participants who changed their minds interacted with AI models in the debriefing phase. 




During the study, only 25\% of participants interacted with AI models, despite the demo video where the features and prediction models available were explained as well as the  time spent exploring the system. We propose two explanations:  (1) participants may have been tense since they were being observed while they worked in the system (2) participants may have adopted a  task-driven behavior over an exploratory behavior when interacting with the system. 
Design implications include understanding where in the UI users can most benefit from the inclusion of content, guidance, or outputs from AI models as a way to tackle recency bias and promote adoption when new AI-supported tasks are added to a complex AI-supported decision-making system.

The proposed method considers RTA and the use of the AI-supported decision-making system as an artifact for reflection about the tasks performed. The rationale of such approach resides in the complexity of the system and number of available AI-supported tasks. In this sense, RTA is recommended to be performed during the debriefing and CTA is optional, under facilitators' discretion.
Some participants did not verbalize their thoughts despite facilitator encouragement. In such cases, the facilitator identified that the cognitive load was higher for those participants and more time was allowed between interventions. For instance, in three such cases, participants responded, \textit{"I'm thinking..."} Moreover, interesting results emerged in the debriefing phase, when these participants had the opportunity to reflect and describe the rationale and insights while exploring and combining multiple AI models.

The retrospective aspect also brought possibilities for identifying multiple task flows and how AI model outputs can be better integrated. The proposed method was explored in a pilot involving four participants and a main study involving 25 participants. Results point to interesting directions for how to better integrate AI models as part of task flows typically performed in a qualitative manner. For instance, participants often began with the visual analytics capabilities and sometimes forgot about the additional AI capabilities. In the debriefing, P7 and P18 mentioned that they forgot about the models and P21 highlighted that he used the visualization because it provides an overview.
Recalling the strategy people often used for the proposed tasks, participants initially performed more in-breadth analysis before the in-depth analysis, including the interaction with AI model outputs.
In some cases, participants interacted with AI models to obtain a second opinion.
This confirmation played an interesting role even when a participant  incorrectly interpreted an AI output to support his decision (representing confirmation bias).

\section{Conclusion}
\label{sec:conclusion}

According to Hutchins \cite{Hutchins2000}, \textit{``It is essential to distinguish the cognitive properties required to manipulate the artifact from the computation that is achieved via the manipulation of the artifact.''} Hence, it is of key importance to differentiate when people are experiencing cognitive load to process the user interface versus the cognitive load they are experiencing as they get insights from consuming content from the user interface. To this end, this work presented \textbf{Retrospective End-user Walkthrough}, a way of performing RTA inspired by \textit{CW with users} with the goal of addressing  emerging needs for evaluating human-AI decision-making systems combining multiple AI models. The proposed method emphasizes the debriefing phase, when RTA is employed and participants are encouraged to revisit tasks performed, interact with the system as a way to reflect on cognitive processes using the system itself as an artifact for the distributed cognition process --including how multiple AI models were combined. This debriefing step of the proposed method supports the qualitative assessment of interpretability of AI models, trust of AI models in decision-making, and planned interactions from the composition of multiple AI models discussed. Moreover, the proposed method allowed identifying interesting effects as identifying recency bias, confirmation bias, trust via transitivity, and trust by projected similarity.

Finally, decision-making systems and human-AI interactions are often considered in silos of one single model and one specific task. In this work, we present a method for qualitatively assessing how multiple AI models are combined in a decision-making system, including models that are more used, how static and AI models can be integrated and how people can benefit from using AI models, whether to have a more focused in-depth analysis or a second AI-based opinion.
Future work is planned to increase the number of participants and to implement requirements identified for bringing AI model outputs to task flows the users commonly perform in a seamless and integrated way. This has implications for HCI practitioners on identifying such entry points for AI model outputs and  for AI experts on identifying novel ways of integrating predictions in multiple tasks and UI elements.
\bibliographystyle{ACM-Reference-Format}
\bibliography{sample-authordraft}










\end{document}